\documentclass[reprint,pra,twocolumn,floatfix]{revtex4-1}
\pdfoutput=1

\usepackage{amsmath}
\usepackage{graphicx}
\usepackage{subfigure}

\newcommand{\ket}[1]{\left| #1 \right>} % for Dirac bras
\begin{document}
\title{Degenerate Bose-Fermi mixtures of rubidium and ytterbium}
\author{V.D. Vaidya}
\email{varunvai@umd.edu}
\affiliation{Joint Quantum Institute, University of Maryland and NIST, College Park, MD 20742}
\author{J. Tiamsuphat}
\author{S.L. Rolston}
\author{J.V. Porto}
\affiliation{Joint Quantum Institute, University of Maryland and NIST, College Park, MD 20742}

\date{\today}

\begin{abstract}
We report the realization of a quantum degenerate mixture of bosonic $^{87}$Rb and fermionic $^{171}$Yb atoms in a hybrid optical dipole trap with a tunable, species-dependent trapping potential. $^{87}$Rb is shown to be a viable refrigerant for the noninteracting $^{171}$Yb atoms, cooling up to $2.4 \times 10^5$ Yb atoms to a temperature of $T/T_F = 0.16(2) $ while simultaneously forming a $^{87}$Rb Bose-Einstein condensate of $3.5 \times 10^5$ atoms. Furthermore we demonstrate our ability to independently tailor the potentials for each species, which paves the way for studying impurities immersed in a Bose gas.
\end{abstract}

\maketitle
\section{Introduction}
Mixtures of quantum degenerate gases provide a platform for conducting a variety of experiments, such as the creation of polar molecules \cite{Carr2009,Jones2006}, the study of impurities in a Bose gas \cite{Klein2007}, and atom interferometry \cite{Hartwig2015}. In particular, there has been recent interest in observing the behavior of polarons on a lattice \cite{Bruderer2008} and the implementation of novel lattice cooling schemes \cite{Griessner2006,Ramos2014} using a Bose-Einstein condensate (BEC) as a bath. Both of these experiments require independent tuning of the trapping potential for each species. While the ability to create a species-dependent potential has been demonstrated in a spin-mixture of $^{87}$Rb \cite{McKay2013,Chen2014,Gadway2010,Gadway2011,Gadway2012}, a $^{23}$Na-$^{6}$Li mixture \cite{Scelle2013} and a $^{87}$Rb-$^{41}$K mixture \cite{Catani2009}, these methods relied on the light shifts from near-resonant light, which led to heating from large photon scattering rates. In the first case a spin-dependent lattice was created using the vector light shift of rubidium at $790$~nm, while in the latter two, a near resonant $670.5$-nm ($789.9$-nm) beam  created a large light shift for lithium (potassium) but a negligible one for sodium (rubidium). Mixtures of alkali-metal and alkaline-earth atoms \cite{Hansen11a,Pasquiou2013,Tassy2007} on the other hand are ideal for such experiments because their atomic spectra differ greatly, allowing one to use light shifts from substantially different wavelengths to create species-dependent traps.

We demonstrate the realization of a degenerate mixture of bosonic $^{87}$Rb and fermionic $^{171}$Yb atoms. Unlike the Li-Yb and Rb-Sr mixtures presented in \cite{Hansen11a,Pasquiou2013}, cooling the Rb-Yb mixture to degeneracy has been challenging due to extremely unfavorable $s$-wave scattering lengths \cite{Borkowski2013} between $^{87}$Rb and most Yb isotopes. $^{172}$Yb and $^{176}$Yb do not form large BECs due to negative self-scattering lengths, and $^{174}$Yb phase separates from $^{87}$Rb well before the onset of degeneracy \cite{Baumer2011} due to a large positive interspecies scattering length. While a degenerate $^{87}$Rb-$^{170}$Yb mixture may be possible, the $s$-wave interaction between the two species is negligible, leaving $^{171}$Yb and $^{173}$Yb as the only two viable candidates for a stable, interacting mixture with $^{87}$Rb.

Our mixture is trapped and evaporated in a species-dependent potential created by a bichromatic optical dipole trap (BIODT) \cite{Onofrio2002,Onofrio2004,Tassy2007}, an ytterbium-blind magnetic quadrupole trap \footnote{The ground state of $^{171}$Yb only has a nuclear spin and a weak magnetic moment, which leaves it largely unaffected by the magnetic quadrupole trap.}  and a rubidium-blind crossed-dipole trap at $423.018$~nm, where the light shift for $^{87}$Rb vanishes \cite{Herold2012}. While this Yb isotope cannot be evaporated by itself due to its small s-wave scattering length of $-3a_0$, $^{87}$Rb serves as a coolant with a favorable interspecies scattering length of $-59a_0$ \cite{Borkowski2013} and we are able to sympathetically cool $2.4 \times 10^5$ Yb atoms down to $T = 0.16T_F$. This represents a tenfold improvement in number over previous methods \cite{Taie2010} to cool $^{171}$Yb. Furthermore we show that this mixture is long lived and not significantly limited by scattering from any dipole beams.

The BIODT consists of co-propagating, overlapped, $1064$-nm (red) and $532$-nm (green) dipole traps. While both beams provide an attractive potential for Yb, the red trap is attractive for Rb while the green is repulsive, allowing for independent control of trap depths for both species. A $423$-nm crossed dipole beam provides longitudinal confinement for ytterbium atoms while the magnetic quadrupole field does the same for Rb atoms, giving us the ability to create species-dependent potentials for efficient sympathetic cooling of Yb atoms.

Our approach, outlined in Fig. \ref{fig:cycle}, is to load laser-cooled nonmagnetic Yb atoms into the green dipole trap and hold them while a Rb magneto-optical trap (MOT) is prepared and transferred into the magnetic quadrupole trap for forced rf evaporation. The two MOTs have to be loaded sequentially since light-assisted collisions in a two-species MOT severely limit its lifetime and size. The repulsive potential created by the green beam prevents hot Rb atoms from heating the colder Yb atoms out of the dipole trap. After RF evaporation of the Rb to the temperature of the Yb cloud, the red beam is turned on, changing the repulsive green potential into an attractive BIODT potential and initiating thermal contact between the two species. Evaporation of Rb in the BIODT is then performed to cool the rubidium and ytterbium clouds to degeneracy.
\begin{figure}[h!]
\includegraphics[scale=1]{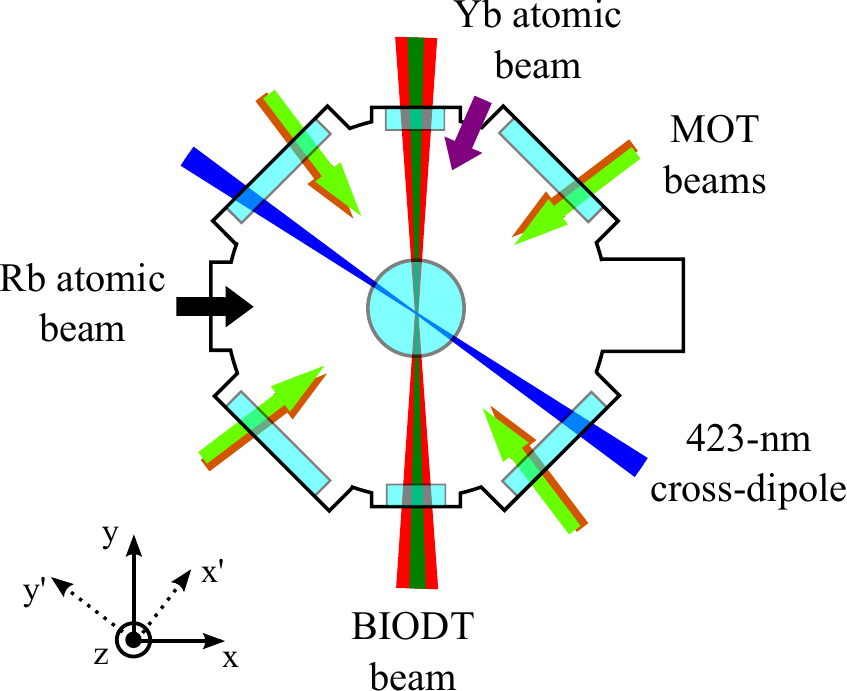}
\caption{\label{fig:traps}(Color online) Experimental setup for producing degenerate mixtures of rubidium and ytterbium. The $27$-$\mu$m BIODT beam intersects the $50$-$\mu$m wide $423$-nm crossed dipole beam at an angle of $56^{\circ}$. MOT beams for both species are retro-reflected through the chamber windows. Gravity is along the z-axis.}
\end{figure}

\section{Apparatus}
The apparatus for creating degenerate mixtures is an adaptation of the Rb BEC system described in \cite{lin09b} with a Yb 2D-MOT addition, similar to the setups described in \cite{tiecke09a,Dorscher2013}. The two-dimensional (2D) MOT uses a field gradient of $5.5$~mT/cm ($55$~G/cm) provided by a pair of permanent magnets and is fed by an Yb oven heated to $410$~$^{\circ}$C. The 2D-MOT beams, detuned by $-50$~MHz from the $^1$S$_0 \rightarrow ^1$P$_1$ transition in Yb, have an intensity of $1.2I_s$ and are retro-reflected through the 2D-MOT beam viewports. A push beam detuned by $+10$~MHz from the $^1$S$_0 \rightarrow ^3$P$_1$ transition accelerates atoms along the axis of the 2D-MOT into the experiment chamber where they are captured and cooled in a three-dimensional (3D) MOT operating on the $^1$S$_0 \rightarrow ^3$P$_1$ line as described in \cite{Kuwamoto1999}.

Using this configuration, we achieve a 3D-MOT loading rate of up to $6 \times 10^7$~s$^{-1}$ for $^{174}$Yb. The loading rate of the 3D-MOT scales with isotope abundance for the bosonic isotopes $^{170}$Yb, $^{172}$Yb, $^{174}$Yb and $^{176}$Yb. However the loading rate for the fermions $^{171}$Yb and $^{173}$Yb is significantly worse than scaling with abundance due to a large differential Zeeman splitting between the ground and excited states \cite{Mukaiyama2003,Dorscher2013}. Two-dimensional MOTs of $^{173}$Yb are further complicated due to an unresolved hyperfine splitting in the $^1$P$_1$ state.  With $^{171}$Yb, we achieve a loading rate of $2.5 \times 10^6$~s$^{-1}$.

The red and green BIODT beams are sent to the experiment over separate optical fibers and focused at the center of the chamber with a waist of $27$~$\mu$m, as inferred from resonant parametric heating measurements of Rb and $^{174}$Yb. Large mode area photonic band-gap fibers ensure a clean TEM$_{00}$ spatial mode for each of the dipole beams and are required to  handle the high $532$-nm power typically used in the experiment. Confinement along the axis of the BIODT is provided by the $423$-nm crossed dipole trap for Yb and a quadrupole magnetic field for Rb. The power in each dipole beam is actively stabilized. 

Atom number and temperature measurements of both species are performed by taking resonant absorption images of each atomic cloud in the $y'z$ and $xy$ planes shown in Fig. \ref{fig:traps}. The Yb cloud is imaged on the broad $^1$S$_0 \rightarrow ^1$P$_1$ transition at $398.9$~nm while the Rb cloud is imaged on the D2 line at $780.2$~nm. Dichroic mirrors placed in the imaging path separate the Rb and Yb images onto different CCD cameras, allowing for simultaneous imaging of both clouds in both $y'z$ and $xy$ planes.

\section{Experimental sequence}
\subsection{Rubidium and ytterbium MOTs}
Since ground-state Yb atoms are insensitive to magnetic fields, we first transfer Yb into the dipole trap. An $8$-s MOT loading stage loads $2 \times 10^7$ Yb atoms into the 3D-MOT with a field gradient of $2.4$~G/cm. The loading rate of the Yb MOT is improved by spectrally broadening the $556$-nm MOT beams from $20$~kHz to $5$~MHz in order to increase the MOT capture velocity. A $200$-ms-long cooling and compression stage reduces the MOT beam linewidth to $20$~kHz, reduces the intensity from $10I_s$ to $1.2I_s$, and increases the field gradient from $2.4$ to $12$~G/cm resulting in a cloud with a density of $3 \times 10^{10}$~cm$^{-3}$ and a temperature of $7$~$\mu$K. We then move the compressed MOT, using uniform magnetic fields, onto the focus of the BIODT before the power in the MOT beams is ramped down, transferring $1.5 \times 10^6$ atoms into the trap at $45$ $\mu$K. At this stage the Yb trapping potential of the BIODT is provided by the $532$-nm beam at $5$~W and the $423$-nm beam at $72$~mW, resulting in a $325$~$\mu$K trap depth. The initial temperature of the Yb cloud is much larger than the $4$~$\mu$K depth of the crossed dipole beam, so its contribution at this stage is negligible.
\begin{figure}
\includegraphics[scale=1]{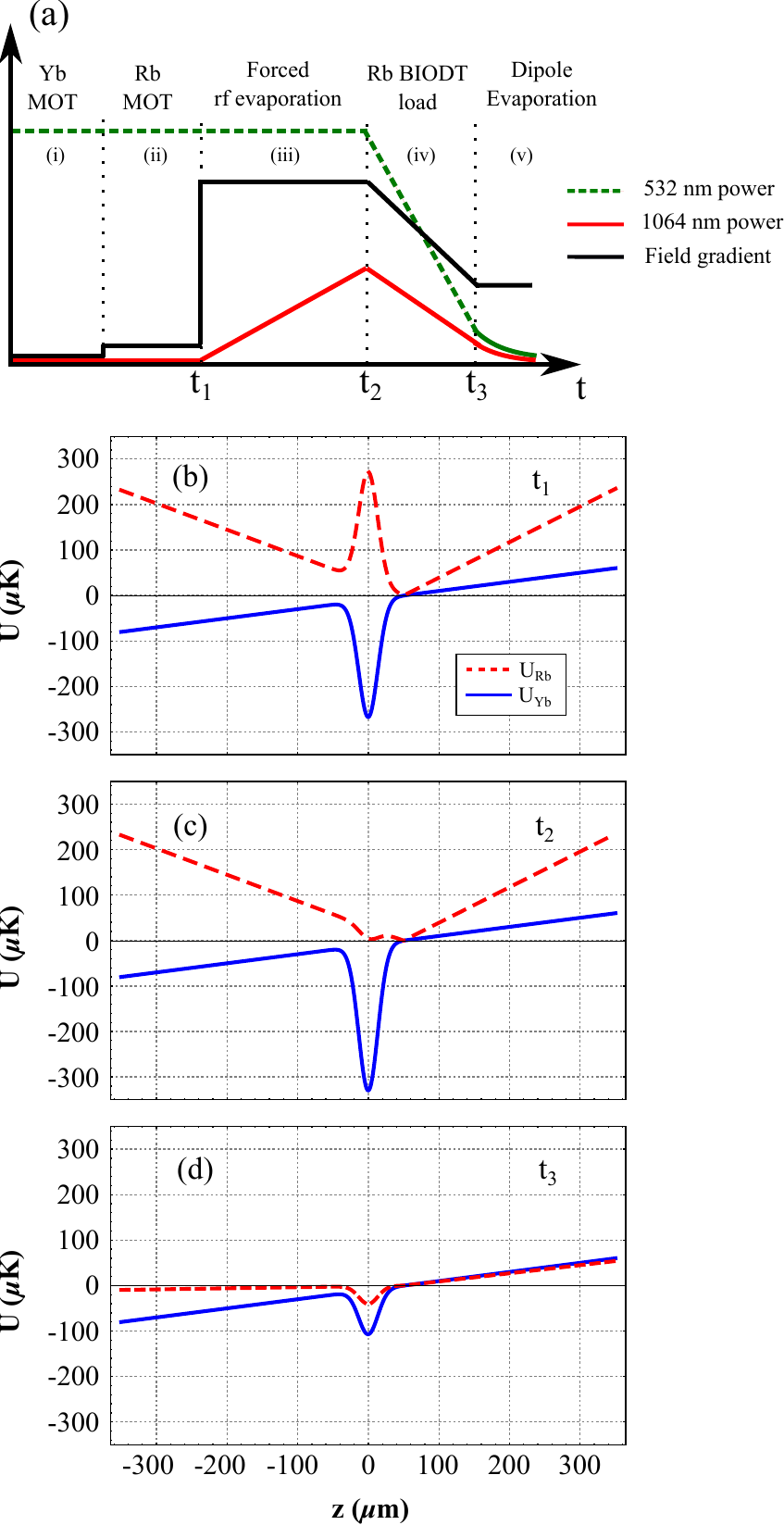}
\caption{\label{fig:cycle}(Color online) (a) Experimental sequence for creating degenerate mixtures of $^{87}$Rb and $^{171}$Yb from steps (i) to (v). (i) $^{171}$Yb MOT and transfer into BIODT, (ii) $^{87}$Rb MOT and transfer into magnetic trap (iii) RF evaporation of $^{87}$Rb in magnetic trap, (iv) transfer of $^{87}$Rb into BIODT accompanied by sympathetic cooling of  $^{171}$Yb and (v) dipole evaporation of $^{87}$Rb to degeneracy. (b)-(d) show plots of the full trapping potential for Rb and Yb during points $t_1$, $t_2$, and $t_3$ respectively.}
\end{figure}

We load the Rb MOT from a Zeeman-slowed atomic beam $2.5$~mm above the BIODT to avoid light-assisted collisions heating Yb atoms out of the trap. After a $6$-s MOT loading stage and a compression and optical pumping stage, $3.5 \times 10^8$ atoms of Rb in the $\ket{F=1,m_F=-1}$ state are transferred into the magnetic quadrupole trap at a field gradient of $192$~G/cm. The field zero of the quadrupole trap is situated $50$~$\mu$m above the BIODT and thermalization between the $105$~$\mu$K Rb and the $45$~$\mu$K Yb clouds is initially suppressed by the large repulsive potential of the green BIODT beam for Rb [Fig.\ref{fig:cycle}(b)].
\subsection{Radiofrequency evaporation and Rb BIODT load}
\begin{figure}
\includegraphics[scale=1]{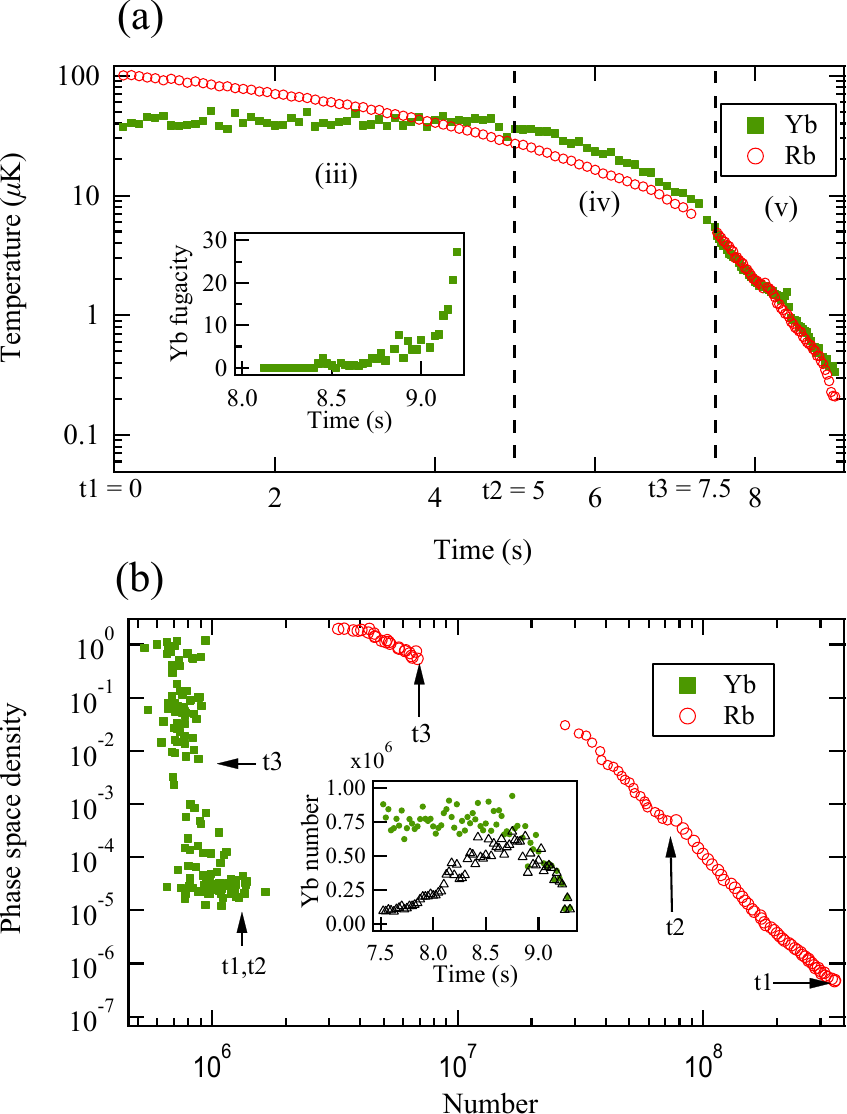}
\caption{\label{fig:data}(Color online) (a) Temperature evolution of Rb and Yb clouds throughout the evaporation procedure. The labeled steps and times correspond to the ones shown in Fig \ref{fig:cycle}. The evolution of the Yb fugacity is shown in the inset. (b) The evolution of the phase space density and number during the evaporation procedure. The steep slope of the Yb data is indicative of efficient sympathetic cooling by Rb. The inset shows the evolution of the Yb number over time during the BIODT evaporation. The total number is given by green circles while the number in the crossed dipole trap is represented by black triangles.}
\end{figure}

The first stage of cooling involves forced RF evaporation of the Rb cloud in the magnetic trap. During this stage [stage (iii) in Fig. \ref{fig:cycle}(a)], we evaporate the Rb cloud using RF from $18$~MHz to $5$~MHz over $5$~s, reducing its temperature to $27$~$\mu$K and increasing its density from $7 \times 10^{10}$~cm$^{-3}$ to $1 \times 10^{13}$~cm$^{-3}$. At the same time we linearly increase the power of the red BIODT beam to $2$~W in order to gradually initiate thermal contact with the Yb cloud towards the end of this stage. Throughout this process, the Yb number and temperature remain constant at $1.5 \times 10^6$ and $45$~$\mu$K (Fig. \ref{fig:data}).

\begin{figure}
\includegraphics[scale=1]{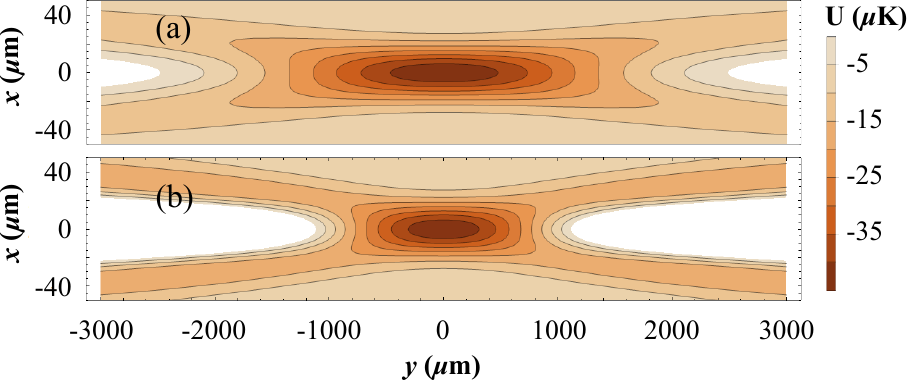}
\caption{\label{fig:CaptureVol}(Color online) The BIODT potential for a $45$~$\mu$K deep Rb trap created using two different powers. The white regions are repulsive. (a) Rb BIODT potential for $0.8$~W and $1.4$~W in the $1064$-nm and $532$-nm beams respectively. (b) Rb BIODT potential for $2.1$~W and $5$~W  in the $1064$- and $532$-nm beams respectively.}
\end{figure}

We then load the Rb cloud into the BIODT by decompressing the magnetic trap to $22$~G/cm while further lowering the RF frequency from $5$~MHz to $1$~MHz over $2.5$~s [stage (iv) in Fig. \ref{fig:cycle}(a)]. The magnetic field gradient of $22$~G/cm provides some levitation against gravity for Rb and confinement along the BIODT with a trap frequency of $8$~Hz. A linear ramp of BIODT powers, as shown in stage (iv) of Fig. \ref{fig:cycle}, increases the overlap between the two species and cools the Yb. Optimal loading of Rb into the BIODT occurs when the green power is reduced to $1.4$~W and the red power to $0.8$~W. We attribute this to a reduced capture volume at high powers, which stems from differing Rayleigh lengths of the $532$- and $1064$-nm BIODT beams when their waists are matched as shown in Fig. \ref{fig:CaptureVol}. The temperature of both species is reduced to $6$~$\mu$K at the end of decompression as shown in Fig. \ref{fig:data}(a) and the Yb cloud suffers minimal loss during this stage. At the end of stage (iv), a Rb cloud containing $7 \times 10^6$ atoms and a Yb cloud containing $8 \times 10^5$ atoms coexist in the BIODT at $6$~$\mu$K.

A Ramsauer-Townsend scattering minimum at $50$~$\mu$K \cite{Frye2014} in the interspecies thermalization cross section causes the $^{171}$Yb temperature to lag behind the $^{87}$Rb temperature during stage (iv). At this temperature, the interspecies cross section drops to $1.2 \times 10^3$~$a_0^2$, well below its low energy s-wave value of $4.4 \times 10^4$~$a_0^2$. However, the thermal distribution of velocities allows us to cool the Yb cloud despite the existence of a scattering minimum, and the two cloud temperatures converge to the same value towards the end of stage (iv).
\subsection{BIODT evaporation to degeneracy}
\begin{figure}
\includegraphics[scale=1]{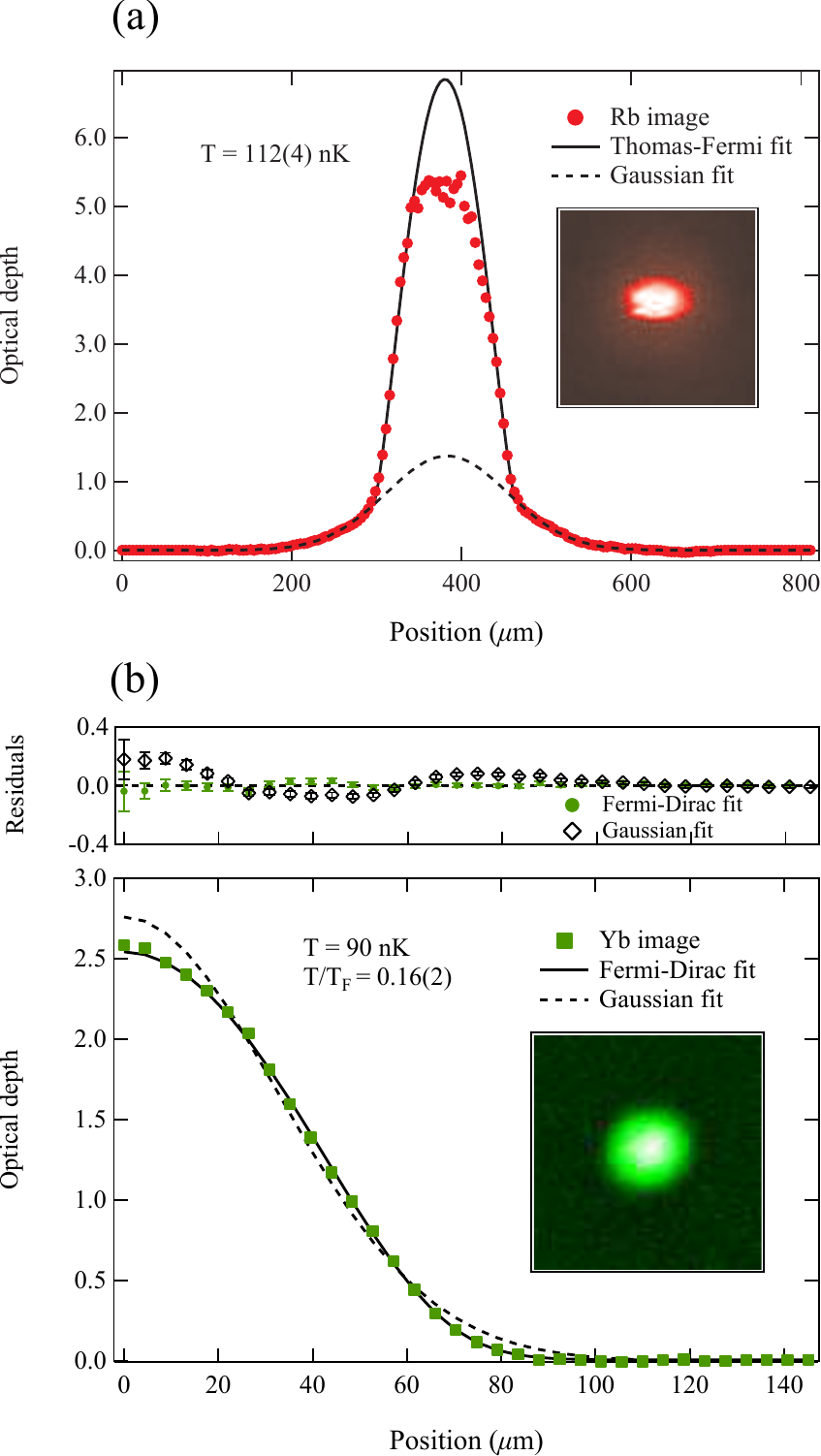}
\caption{\label{fig:FDfits}(Color online) (a) A $22$-ms time-of-flight (TOF) image of a condensed Rb cloud of $5 \times 10^5$ atoms at $112$~nK, as extracted from a Gaussian fit to the wings of the image. The plot shows a slice of the image across 10 pixels, along with the fits. The bimodal fit is performed only on points with an OD below 3 in order to avoid artifacts from saturation of the probe beam. (b) A $12$-ms TOF image of a degenerate Yb cloud at $90$~nK with $2.4 \times 10^5$ atoms. The plot shows azimuthal averages of the image along with the 2D fits performed on the image. At this temperature the Rb cloud forms a nearly pure BEC, without any discernible thermal wings.}
\end{figure}

We perform the final stage of evaporation [stage (v) in Figs. \ref{fig:cycle} and \ref{fig:data}] with an exponential ramp of both BIODT powers with a time constant $\tau = 0.5$~s over a period of $2$~s. Towards the end of this stage the Rb cloud condenses to form a BEC while Yb reaches Fermi degeneracy. The Rb images show a strong bimodal distribution characteristic of a BEC, while the Yb cloud shows significant deviations from a Gaussian distribution (Fig. \ref{fig:FDfits}). After $1.7$~s of evaporation in the BIODT, the Yb cloud reaches a temperature of $T = 0.62(8)T_F = 220$~nK with $4.2 \times 10^5$ atoms while the Rb cloud condenses to form a BEC of $3.5 \times 10^5$ atoms. Evaporating for $2$~s cools the Yb cloud to $T = 0.16(2)T_F = 90$~nK with $2.4 \times 10^5$ atoms, at the cost of reducing the Rb BEC number to $1.1 \times 10^5$. This is over an order of magnitude improvement over previous methods to cool $^{171}$Yb \cite{Taie2010} which resulted in $8 \times 10^3$ atoms at $T = 0.46T_F$. The uncertainties in our values for $T/T_F$ are dominated by noise in the number of Yb atoms and not by our temperature measurement.

Near degeneracy, we extract the number and temperature of the $^{171}$Yb cloud from a two-dimensional fit of a Fermi-Dirac distribution,
\begin{equation}
n(x,y) = B - A\frac{\mathrm{Li}_2\left[-z \exp \left(\frac{x^2}{w_x^2}+\frac{y^2}{w_y^2}\right)\right]}{\mathrm{Li_2} \left(-z\right)}
\end{equation}
to the absorption image, with fit parameters $A$, $B$, $z$, $w_x$, and $w_y$. $\mathrm{Li}_2$ is a polylogarithm of order 2 and $z = e^{\mu/{k_B T}}$ is the fugacity that determines the deviation of the Fermi-Dirac distribution from a thermal Gaussian. For a thermal gas at $T \gg T_F$, $z = 0$ while in the limit $T \ll T_F$ the fugacity diverges. We extract the temperature of the degenerate Fermi gas from $w_x$ and $w_y$. The number of atoms is determined using two methods: through direct integration of the column density, and indirectly from the fugacity and temperature using the relation
\begin{equation}
\frac{T}{T_F \left(\omega,N \right)} = \frac{1}{\mathrm{ln}(z)}\mathrm{,}
\end{equation}
which holds in the limit $T \ll T_F$. The former method gives a value of $2.4 \times 10^5$ atoms while the latter gives a value of $3.1 \times 10^5$, using calculated Yb trap frequencies, $\left( \omega_x, \omega_y, \omega_z \right) = (150,140,75)$~Hz and assuming an unpolarized Fermi gas \footnote{The extremely small energy difference between the $m=1/2$ and $m=-1/2$ sublevels of $^{171}$Yb guarantees our magnetic field variations will be  non-adiabatic, leading to an unpolarized gas.}.

The inset in Fig. \ref{fig:data}(a) shows the fugacity while the inset in Fig. \ref{fig:data}(b) shows the evolution Yb number during the final stage of evaporation. The fugacity rises rapidly during the last second of the evaporation as the cloud becomes degenerate but the number, which does not change significantly during the early stages of evaporation, drops as the trap becomes too shallow to support Yb against gravity.

\subsection{Trap lifetimes}
\begin{figure}
\includegraphics[scale=1]{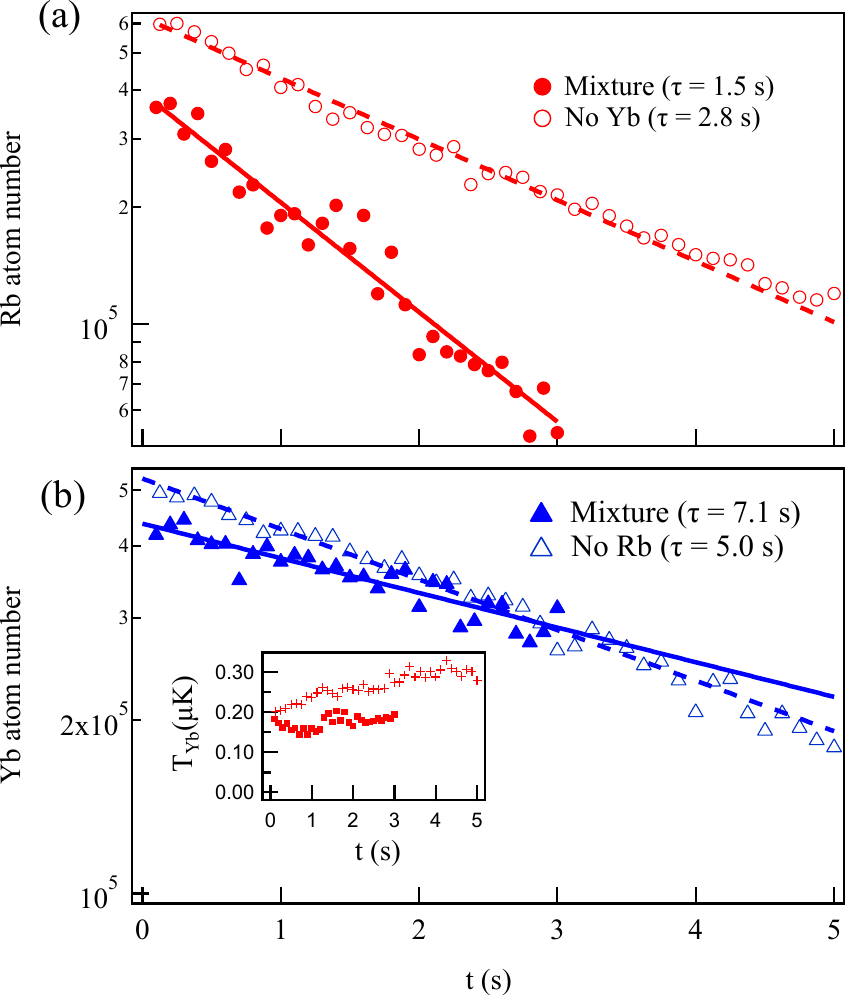}
\caption{\label{fig:lifetimes}(Color online) (a) Lifetime of the Rb BEC with and without the degenerate Yb gas present. (b) Lifetime of the Yb gas with and without the Rb BEC present. The inset in (b) shows the evolution of the Yb temperature in the presence (circles) and absence (crosses) of the Rb BEC. The lines in both plots are fits of an exponential decay to the data.}
\end{figure}

The lifetime of the degenerate mixture is limited by photon scattering of Yb atoms from the $423$~nm beam with calculated rate of $0.4$~s$^{-1}$. We measure the lifetimes of both degenerate clouds under three different conditions shown in Fig. \ref{fig:lifetimes}: Rb in the absence of Yb, Yb in the absence of Rb, and a degenerate mixture. Scattering of $423$-nm photons heats the Yb cloud in the absence of Rb, as shown in the inset in Fig. \ref{fig:lifetimes}(b). In the presence of the Rb BEC, this heating rate is reduced and the Rb lifetime is reduced from $2.8$~s to $1.5$~s  [Fig. \ref{fig:lifetimes}(a)] as the Rb cloud evaporates to keep the Yb temperature fixed. The Rb cloud has a peak density of $2.8(5) \times 10^{14}$~cm$^{-3}$ and its lifetime in the absence of Yb is limited by three-body recombination. Using a $^{87}$Rb three-body rate constant of $K_3 = 5.8 \times 10^{-30}$~cm$^6$s$^{-1}$ \cite{Burt1997}, we calculate a three body rate of $0.45$~s$^{-1}$, in approximate agreement with our measured lifetime.

\section{Independent control of Rb and Yb clouds}
The flexible nature of our trap allows us to control overlap between the two species, and independently address either Rb or Yb atoms. The nonmagnetic Yb atoms are unaffected by the magnetic trap used to confine the Rb cloud along the BIODT beam. Similarly the Rb cloud does not see the $423$-nm crossed dipole beam that provides longitudinal confinement for the Yb gas.
\begin{figure}[t!]
\includegraphics[scale=1]{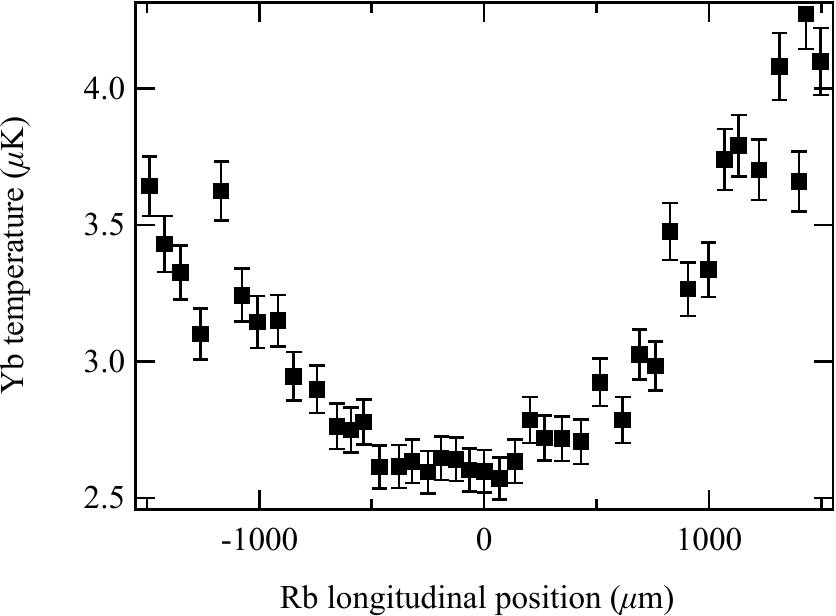}
\caption{\label{fig:overlap} Yb temperature after BIODT evaporation, as a function of the position of the Rb cloud. A minimum is observed at $2.5$~$\mu$K corresponding to the optimal longitudinal overlap between the two clouds.}
\end{figure}

The overlap between the two clouds can be controlled by using a uniform magnetic field to shift the position of the magnetic trap along the BIODT. Before BIODT evaporation, the magnetic Rb atoms are moved along the dipole trap while the position of the nonmagnetic Yb atoms is left unchanged. We use the Yb temperature after $0.5$~s of BIODT evaporation as an indicator of overlap between the two species, as the noninteracting Yb atoms cannot be efficiently evaporated. This temperature is shown in Fig. \ref{fig:overlap} as a function of the Rb cloud position, with a clear minimum in the Yb temperature indicating optimal longitudinal overlap of the two clouds.

\begin{figure}[h!]
\includegraphics[scale=1]{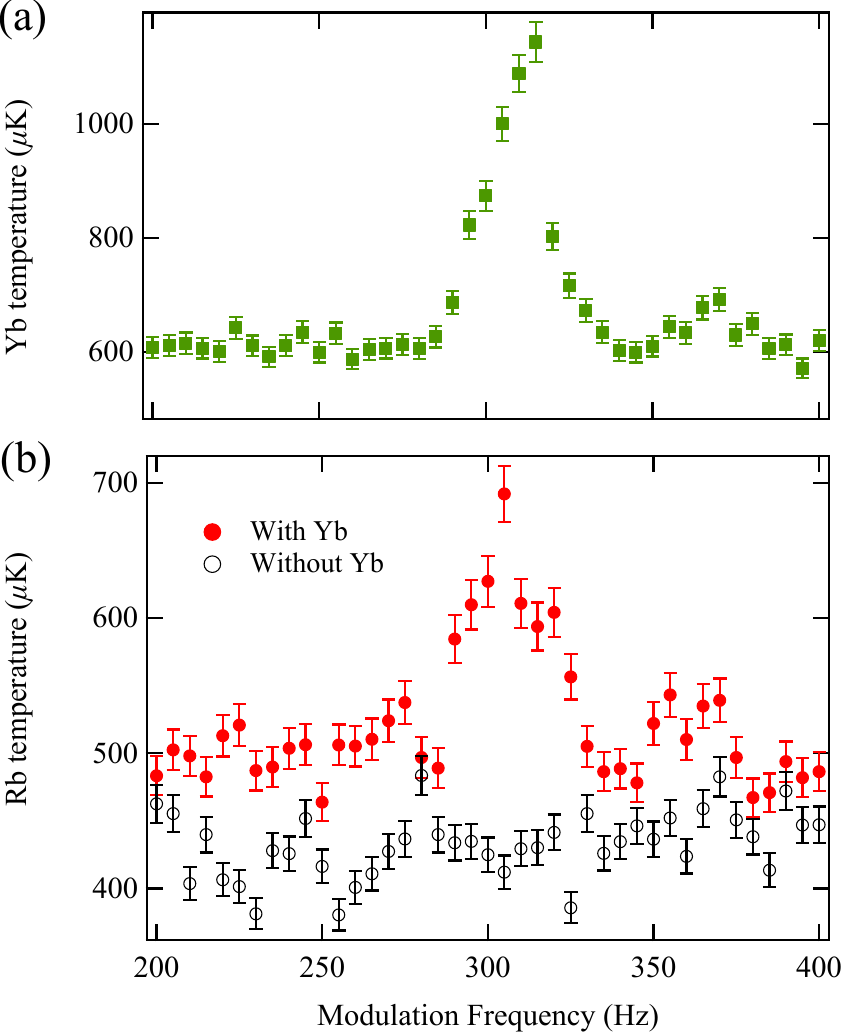}
\caption{\label{fig:parametric}(Color online) (a) Yb temperature as a function of $423$~nm modulation frequency, showing a resonance at $310$~Hz. (b) Rb temperature as a function of modulation frequency in the presence, (red filled circles) and absence (black empty circles) of Yb. The Rb cloud is heated only in the presence of Yb confirming that both species are overlapped and that the $423$~nm beam does not affect Rb.}
\end{figure}

Independent manipulation of the Yb cloud can be achieved through the $423$-nm crossed dipole beam. To demonstrate this, we resonantly heat the Yb by modulating the intensity of the crossed dipole beam at the trap frequency (Fig. \ref{fig:parametric}(a)). This in turn heats up the Rb, resulting in a rise in its temperature as shown in Fig. \ref{fig:parametric}(b). When this experiment is repeated in the absence of Yb, we observe no heating of the Rb cloud, confirming that the two clouds are overlapped and that the crossed dipole beam only affects Yb.
\vspace{0.7cm}
\section{Conclusions and outlook}
We have created a degenerate Bose-Fermi mixture of $^{87}$Rb and $^{171}$Yb in a species-dependent potential. Our ability to independently tune trap potentials allows us to sympathetically cool Yb with minimal loss, allowing us to create large degenerate Fermi gases. We further demonstrate that the degenerate mixtures are stable and long lived, with the loss rate limited by slow photon scattering from the $423$-nm crossed dipole trap.

A species-selective optical lattice created by the $423$-nm laser will allow us to realize several lattice cooling schemes \cite{Griessner2006,Ramos2014}. Furthermore, our method of creating degenerate mixtures is extremely flexible and can be extended to other alkali metals such as sodium or cesium, providing a route for a variety of other degenerate alkali-metal--alkaline-earth mixtures.
\begin{acknowledgments}
The authors thank C. D. Herold and X. Li for their contributions in building the apparatus. We acknowledge useful discussions with M. D. Frye and J. M. Hutson. This work was supported by NSF-PHY1104472 and ARO Grant W911NF-07-1-0493, with funds from the DARPA Optical Lattice Emulator program.
\end{acknowledgments}

\bibliography{DegenerateRbYb}
\end{document}